\documentclass[aps,preprint,tightenlines,floatfix,superscriptaddress,nofootinbib]{revtex4}

\usepackage{graphicx}
\usepackage{dcolumn}
\def\beq{\begin{equation}}
\def\eeq{\end{equation}}
\def\bea{\begin{eqnarray}}
\def\eea{\end{eqnarray}}
\def\nn{\nonumber}
\def\sss{\scriptstyle}

\def\roughly#1{\mathrel{\raise.3ex\hbox
{$#1$\kern-.75em\lower1ex\hbox{$\sim$}}}}
\def\lsim{\roughly<}
\def\gsim{\roughly>}

\begin{document}
\bibliographystyle{apsrev}

\preprint{\vbox {\hbox{UdeM-GPP-TH-06-148}}}

\vspace*{2cm}

\title{\boldmath CP Violation in Supersymmetric Theories: \\
${\tilde t}_2 \to {\tilde t}_1 HH$,
${\tilde t}_2 \to {\tilde t}_1 ZZ$,
${\tilde t}_2 \to {\tilde t}_1 W^+ W^-$,
${\tilde t}_2 \to {\tilde t}_1 ZH$}

\def\umontreal{\affiliation{\it Physique des Particules, Universit\'e
de Montr\'eal, \\ C.P. 6128, succ. centre-ville, Montr\'eal, QC,
Canada H3C 3J7}}
\def\tayloru{\affiliation{\it Physics Department, Taylor University, \\
236 West Reade Ave., Upland, IN 46989, USA}}

\umontreal
\tayloru

\author{Alejandro Szynkman}
\email{szynkman@lps.umontreal.ca}
\umontreal

\author{Ken Kiers}
\email{knkiers@taylor.edu}
\tayloru

\author{David London}
\email{london@lps.umontreal.ca}
\umontreal

\date{\today}

\begin{abstract}
We study the decays ${\tilde t}_2 \to {\tilde t}_1 HH$, ${\tilde t}_2
\to {\tilde t}_1 ZZ$, ${\tilde t}_2 \to {\tilde t}_1 W^+ W^-$ and
${\tilde t}_2 \to {\tilde t}_1 ZH$, with an eye towards measuring the
CP-violating supersymmetric parameters contained in these processes. We
find that ${\tilde t}_2 \to {\tilde t}_1 HH$ tends to have the largest
CP asymmetry and width, and is perhaps the most favourable
experimentally. These decays are sensitive primarily to $\phi_{\sss
A_t}$, the phase of the trilinear coupling $A_t$.
\end{abstract}


\maketitle

\newpage
\setcounter{page}{1}

Supersymmetry (SUSY) is widely thought to be the physics that lies
beyond the standard model (SM). SUSY theories typically contain many
new parameters, some of which are complex, and hence violate CP.
Should SUSY be found experimentally, we will want to find the values
of all of these parameters. In particular, it will be important to
measure the CP-violating SUSY phases.

CP-violating SUSY effects have been studied extensively in meson
mixing \cite{mixing}, in CP violation related to the $B$-meson system
\cite{B-meson} and in electric dipole moments (EDMs) \cite{edms}. In
particular, EDMs provide quite stringent constraints on the low-energy
CP-violating SUSY phases of the superparticle couplings. For example,
if sfermion masses are taken to be of order the weak scale and complex
SUSY parameters have phases of order unity, theoretical predictions for
EDMs are not generically in agreement with experimental
limits. Nevertheless, the so-called SUSY CP problem can be avoided in
several SUSY scenarios \cite{scenarios} as calculations of EDMs are
highly model-dependent. In this paper, all SUSY parameters used as
inputs in the numerical simulations are assumed not to violate EDM
constraints. The approach adopted here is that our observables may
offer an $independent$ measurement of relevant CP-odd SUSY parameters.

In a recent paper \cite{SUSYCP}, we showed that the decay ${\tilde
t}_2 \to {\tilde t}_1 \tau^- \tau^+$ is particularly sensitive to
$\phi_{\sss A_t}$, the phase of the trilinear coupling $A_t$. (The
``stops'' ${\tilde t}_1$ and ${\tilde t}_2$ are the two mass
eigenstates of the scalar superpartners of the top quark, with
$m_{{\tilde t}_2} > m_{{\tilde t}_1}$.) In the present paper, we
examine how some other decay processes depend on the SUSY phases.
Throughout we assume that SUSY has been discovered, and that the
CP-conserving parameters (e.g.\ masses of SUSY particles) are known
independently. 

All effects which violate CP require the interference of (at least)
two amplitudes. For a given decay, there are two types of CP-violating
signals. Direct CP asymmetries (spin-independent or spin-dependent)
are proportional to $\sin\delta$, where $\delta$ is the relative
strong (CP-even) phase between the interfering amplitudes.
Triple-product (TP) asymmetries take the form $\vec v_1 \cdot (\vec
v_2 \times \vec v_3)$ (each $v_i$ is a spin or momentum), and are
proportional to $\cos\delta$. One can therefore have a nonzero TP even
if $\delta = 0$.  It is also possible to have a nonzero TP with only a
single decay amplitude if, for example, the intermediate particle has
both scalar and pseudoscalar couplings~\cite{SUSYCP,ValWang}.

Strong phases can be generated in one of two ways. First, one can have
the exchange of gluons between the particles involved in the decay,
leading to QCD-based strong phases. Unfortunately, we do not know how
to calculate the strong phases in this case. Alternatively, the strong
phases can be generated by the (known) widths of the intermediate
particles in the decay. Given that we want to {\it measure} the SUSY
CP phases, and not simply detect the presence of CP violation, we must
consider decays in which the strong phases are generated in the second
way. Thus, the decay processes cannot contain too many particles which
couple to gluons.

It is therefore quite natural to consider the decays ${\tilde t}_2 \to
{\tilde t}_1 HH$, ${\tilde t}_2 \to {\tilde t}_1 ZZ$, ${\tilde t}_2
\to {\tilde t}_1 W^+ W^-$ and ${\tilde t}_2 \to {\tilde t}_1 ZH$,
where $H$ is a neutral Higgs boson. There are several points here
which we should note in relation to these processes. First, SUSY
theories involve two Higgs doublets which contain (in the gauge basis)
two neutral scalars and one pseudoscalar. In the mass basis, these
particles mix and one obtains three mass eigenstates $H_1$, $H_2$ and
$H_3$. All four decays receive contributions from several diagrams,
including those with an intermediate $H_i$ ($i=1,2,3$). The amplitudes
are typically dominated by those diagrams in which an intermediate
particle can go on shell. In SUSY theories, the lightest mass is
$m_{H_1} = O(100)$ GeV and we therefore take the final-state $H$ to be
$H_1$. When it appears as an internal line in a diagram, the $H_1$ is
too light to decay (on shell) to $H_1 H_1$, $ZZ$, $W^+W^-$ or $ZH_1$.
Nevertheless, we shall retain such diagrams since they can in principle
give non-negligible contributions.

Second, triple products are due to terms of the form ${\rm Tr}
[\gamma_\alpha \gamma_\beta \gamma_\rho \gamma_\sigma \gamma_5]$ in
the square of the amplitude. Since no fermions are involved in the
decay processes, no CP-violating TP's can arise here, and we have only
direct CP asymmetries. The final-state particles do not couple to
gluons and any exchange of gluons between ${\tilde t}_2$ and ${\tilde
t}_1$ only serves to renormalize the couplings of the stops. Thus, the
strong phase arises only due to the widths of the intermediate
particles $H_2$ and $H_3$. For a given set of SUSY parameters, the
widths $\Gamma_2$ and $\Gamma_3$ are calculable (as are the
`off-diagonal widths' associated with transitions $H_i\leftrightarrow
H_j$~\cite{widths}). Thus, the measurement of CP violation in these
decays (due to direct CP asymmetries) will allow us to extract and/or
constrain the SUSY parameters, including the CP-violating SUSY phases.

Third, there can be significant interference between the two decay
amplitudes only if the masses of $H_2$ and $H_3$ are similar.
Fortunately, it is relatively common in SUSY theories that $m_{H_2}
\simeq m_{H_3}$~\cite{higgs_tree}. Since it is assumed that the masses
are known, we will know beforehand whether or not CP violation is
likely in these decays.

Finally, we make a comment regarding our previous work, which examined
CP asymmetries in ${\tilde t}_2 \to {\tilde t}_1 \tau^-\tau^+$
\cite{SUSYCP}. The process ${\tilde t}_2 \to {\tilde t}_1
\tau^-\tau^+$ is interesting since it is theoretically clean (no
strong phases from gluons) and it can have large CP asymmetries while
simultaneously having a moderately large branching fraction. This
process is particularly attractive if the intermediate Higgs bosons
are too light to decay to heavier final states, such as those
considered in the present work ($W^+W^-, H_1H_1$, etc.).
Nevertheless, as noted in Ref.~\cite{SUSYCP}, the simple rate
asymmetry for ${\tilde t}_2 \to {\tilde t}_1 \tau^-\tau^+$ is
extremely small. Thus, sizeable CP asymmetries are only expected in
cases where the spin of one or both of the final-state leptons is
measured. The processes considered in the present work have an
advantage over ${\tilde t}_2 \to {\tilde t}_1 \tau^-\tau^+$ in that
they do not require the measurement of any spins -- the regular rate
asymmetries can have relatively large values.

\begin{figure}[t]
\begin{center}
\resizebox{4in}{!}{\includegraphics*{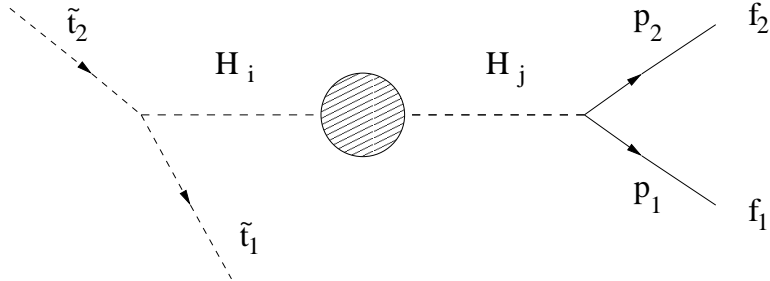}}
\caption{Feynman diagram for the decay $\tilde{t}_2^-\to
\tilde{t}_1^-f_1f_2$ with $f_1f_2=H_1H_1,ZZ,W^+W^-$ and $ZH_1$.  The Higgs
propagator has off-diagonal terms, so transitions $H_i\to H_j$ are
allowed.}
\label{fig:feyn_diag}
\end{center}
\end{figure}

\begin{figure}[t]
\begin{center}
\resizebox{5in}{!}{\includegraphics*{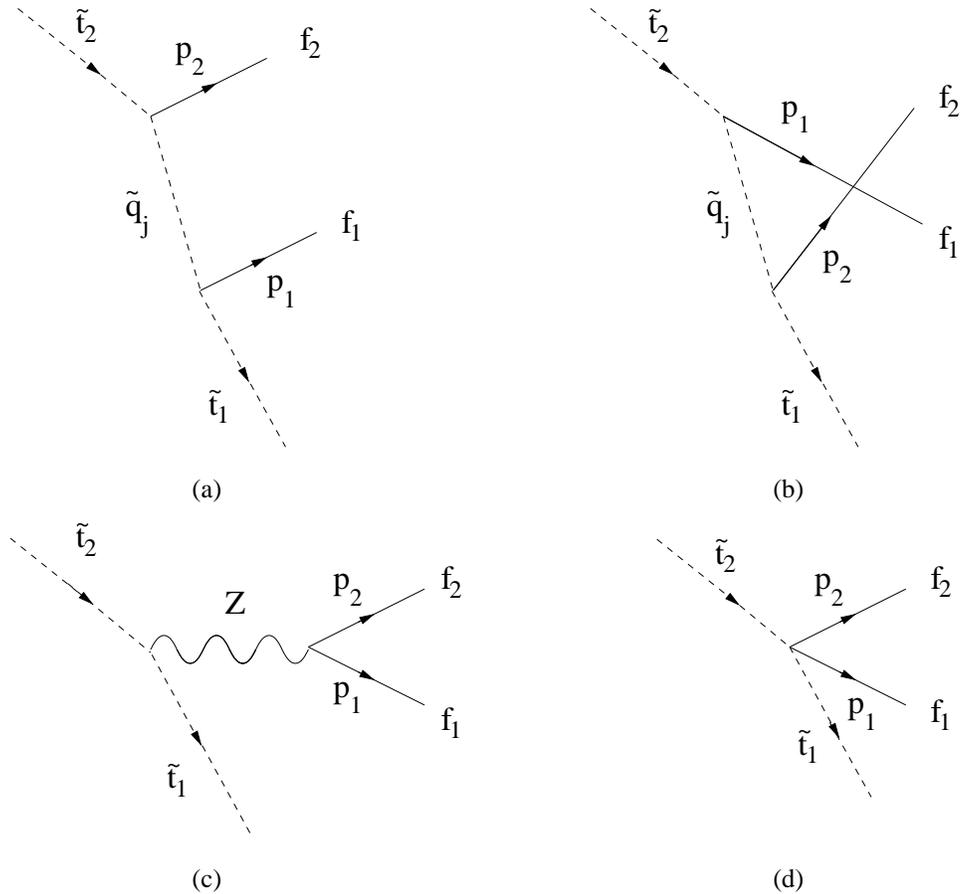}}
\caption{Feynman diagrams that can contribute to the decays $\tilde{t}_2^-\to
\tilde{t}_1^-f_1f_2$ with $f_1f_2=H_1H_1,ZZ,W^+W^-$ and $ZH_1$.  Not all diagrams
contribute to each process.}
\label{fig:feyn_diag2}
\end{center}
\end{figure}

There are five classes of diagrams that contribute to the four
processes under consideration, although not every class of diagram
contributes to each process.  The first class is shown in
Fig.~\ref{fig:feyn_diag}.  In this case the heavier stop decays to the
lighter stop and emits a Higgs boson.  The Higgs then decays to the
final state $f_1f_2$ (where $f_1f_2=H_1H_1$, etc.)  This process can
proceed resonantly if the intermediate Higgs boson(s) can go on shell.
Figure~\ref{fig:feyn_diag2} shows the other four possibilities.  In
diagram (a) the stop decays to particle $f_2$ and a squark
$\tilde{q}_j$ (either a stop or a sbottom), which subsequently decays
into $f_1$ and the lighter stop.  Diagram (b) is a crossed version of
diagram (a).  Diagram (c) is similar to that shown in
Fig.~\ref{fig:feyn_diag}, but with the intermediate Higgs replaced by
a $Z$.  Finally, diagram (d) shows the contribution due to a possible
quadrilinear vertex.

Of all the diagrams shown in Fig.~\ref{fig:feyn_diag2}, there is only
one case in which the process can proceed resonantly.  This occurs for
the decay ${\tilde t}_2^- \to {\tilde t}_1^- W^+ W^-$ ($f_1=W^+$ and
$f_2=W^-$), in which case the intermediate squark in diagram (a) is a
sbottom.  When considering ${\tilde t}_2 \to {\tilde t}_1 W^+ W^-$
in our numerical work we choose parameters in such a way
that the sbottom(s) cannot go on-shell.  Thus, in all cases the
diagrams in Fig.~\ref{fig:feyn_diag2} contribute non-resonantly.
While one might be tempted to ignore these diagrams compared to the
resonant Higgs contributions in Fig.~\ref{fig:feyn_diag}, we have
found that the non-resonant diagrams can sometimes give non-negligible
contributions, so we include them in our calculation.  We will discuss
this point further below.


We turn now to a calculation of the width and rate asymmetry for the
process ${\tilde t}_2 \to {\tilde t}_1 f_1f_2$, with
$f_1f_2=H_1H_1,ZZ,W^+W^-$ and $ZH_1$.  We define two invariant masses
as follows:
\begin{eqnarray}
  M^2 & = & \left(p_1+p_2\right)^2 \; ,\\
  \rho^2 & = & \left(p_1+p_{\tilde{t}_1}\right)^2 \; ,
\end{eqnarray}
where $p_{1,2}$ are the four-momenta of $f_1$ and $f_2$, respectively, and
$p_{\tilde{t}_1}$ is the four-momentum of the ${\tilde t}_1$.  All dot
products that arise in the calculation may be written in terms of
$M^2$, $\rho^2$ and the various particle masses.

As noted above, there are no TP's in this process.  In principle there
are polarization-dependent CP-violating observables similar to the
single-spin CP asymmetry defined in Ref.~\cite{SUSYCP}.  It turns out
that, in the limit in which one can neglect the non-Higgs-mediated
diagrams, such polarization-dependent observables are sensitive to the
same combinations of underlying SUSY parameters as are the rate
asymmetries.  Thus, in this limit, one gains no new information by
measuring polarizations.  In principle the cross terms of resonant and
non-resonant amplitudes can give new contributions to
polarization-dependent CP-odd observables, but such contributions are
expected to be suppressed.  Given the difficulty of the related
measurements, and the possible suppression, we will ignore such
observables and sum over the polarizations of the final-state
particles.

Let us consider first the ``processes'' ${\tilde t}_2^- \to {\tilde
t}_1^- f_1f_2$ (as opposed to the ``anti-processes,'' involving the
decay of a $\tilde{t}_2^+$, which will be considered in a moment).
(Note that the indices $\pm$ associated with the ${\tilde t}$'s
indicate that they have charge $\pm 2/3$.) The Higgs-exchange diagrams shown in
Fig.~\ref{fig:feyn_diag} contribute for all four final states.  The
following are some technical details for each process regarding the
diagrams shown in Fig.~\ref{fig:feyn_diag2}.
\begin{enumerate}

\item ${\tilde t}_2^- \to {\tilde t}_1^- H_1H_1$: Diagrams (a), (b)
and (d) in Fig.~\ref{fig:feyn_diag2} contribute.  (The $H_1$-$H_1$-$Z$
coupling is zero.)  The intermediate $\tilde{q}_j$ are stops.  These
stops cannot go on shell.

\item ${\tilde t}_2^- \to {\tilde t}_1^- ZZ$: Diagrams (a), (b) and
(d) contribute.  The intermediate $\tilde{q}_j$ in diagrams (a) and
(b) are stops; they cannot go on shell.

\item ${\tilde t}_2^- \to {\tilde t}_1^- W^+ W^-$: Diagrams (a), (c)
and (d) contribute ($f_1=W^+$ and $f_2=W^-$).  The intermediate
squarks in diagram (a) are sbottoms. In principle the sbottoms could
go on shell, although in practice we choose parameters such that this
does not occur\footnote{If one or both sbottoms go on shell the
calculation becomes more complicated. In particular, the asymmetry
would depend on the total widths of the sbottoms. Furthermore, gluon
exchange between ${\tilde t}_2$ and ${\tilde t}_1$ in diagram (a)
could give rise to additional strong phases. Even if one were to
include the on-shell sbottoms diagrams, the resulting asymmetry would
tend to decrease. The sbottom diagrams are unlikely to interfere with
each other, since sbottom masses are generically not close to each
other. Also, since sbottoms do not proceed via the $s$-channel, the
on-shell interference between Higgs bosons and sbottoms is restricted
to a very small region in the Dalitz plot.}. 

\item ${\tilde t}_2^- \to {\tilde t}_1^- ZH_1$: Diagrams (a), (b) and (c)
contribute.  The intermediate top squarks in diagrams (a) and (b)
cannot go on shell.

\end{enumerate}

The amplitudes for the four processes may be written in the following
manner:
\begin{eqnarray}
  {\cal A}_{HH} &=& B_{HH} \; , 
    \label{eq:ahh}\\
  {\cal A}_{ZZ} &=& \left(B_{ZZ}\, g^{\mu\nu}+C_{ZZ}\, p_2^\mu p_{\tilde{t}_1}^\nu +
       D_{ZZ}\, p_{\tilde{t}_1}^\mu p_1^\nu+E_{ZZ}\, p_{\tilde{t}_1}^\mu p_{\tilde{t}_1}^\nu  
          \right)\epsilon_\mu^{\lambda_1 *}\epsilon_\nu^{\lambda_2 *} \; , \\
  {\cal A}_{WW} &=& \left(B_{WW}\, g^{\mu\nu}+C_{WW}\, p_2^\mu p_{\tilde{t}_1}^\nu +
       D_{WW}\, p_{\tilde{t}_1}^\mu p_1^\nu+E_{WW}\, p_{\tilde{t}_1}^\mu p_{\tilde{t}_1}^\nu  
          \right)\epsilon_\mu^{\lambda_1 *}\epsilon_\nu^{\lambda_2 *} \; , \\
  {\cal A}_{ZH} &=& \left(B_{ZH}\, p_2^\mu+C_{ZH}\, p_{\tilde{t}_1}^\mu\right)
          \epsilon_\mu^{\lambda_1 *} \; ,
     \label{eq:azh}
\end{eqnarray}
where the $\epsilon_{\mu,\nu}^{\lambda_{1,2} *}$ are the polarization
tensors for the final-state vector mesons.  The parameters
$B_{f_1f_2},\ldots, E_{f_1f_2}$ are functions of the various coupling
constants and mixing matrices.  In each case, $B_{f_1f_2}$ contains
the Higgs-mediated contributions, as well as one or more other
contributions coming from diagram(s) in Fig.~\ref{fig:feyn_diag2};
that is,
\begin{eqnarray}
  B_{f_1f_2} = B_{f_1f_2}^{\mbox{\scriptsize{Higgs}}}+\delta B_{f_1f_2} \; , 
\end{eqnarray}
where $\delta B_{f_1f_2}$ denotes the non-Higgs-mediated
contributions.  Expressions for $\delta B_{f_1f_2}$, $C_{f_1f_2}$,
$D_{f_1f_2}$ and $E_{f_1f_2}$ may be found in the Appendix.  

We include here the expressions for the Higgs-mediated contributions,
$B_{f_1f_2}^{\mbox{\scriptsize{Higgs}}}$, since they are of the most
interest to us (they tend to dominate, and they are also the source of
the required strong phases).  These contributions are given by,
\bea
     B_{HH}^{\mbox{\scriptsize{Higgs}}} & = & - v^2 \sum_{i,j} 
        g_{\sss H_i\tilde{t}_2^*\tilde{t}_1}D_{ij}(M^2)
        g_{\sss H_jH_1H_1} \eta_j\; ,
	\label{eq:bhh} \\
     B_{ZZ}^{\mbox{\scriptsize{Higgs}}} & = & -\frac{vgm_W}{\cos^2\theta_W} \sum_{i,j} 
        g_{\sss H_i\tilde{t}_2^*\tilde{t}_1}D_{ij}(M^2)
        g_{\sss H_jVV} \; ,
        \label{eq:bzz}\\
     B_{WW}^{\mbox{\scriptsize{Higgs}}} & = & -v gm_W\sum_{i,j} 
        g_{\sss H_i\tilde{t}_2^*\tilde{t}_1}D_{ij}(M^2)
        g_{\sss H_jVV} \; ,
	\label{eq:bww}\\
     B_{ZH}^{\mbox{\scriptsize{Higgs}}} & = & \frac{ivg}{\cos\theta_W} \sum_{i,j} 
        g_{\sss H_i\tilde{t}_2^*\tilde{t}_1}D_{ij}(M^2)
        g_{\sss H_1H_jZ} \; ,
	\label{eq:bzh}
\eea
where we have included factors $\eta_1=6$ and $\eta_{2,3}=2$ in order
to correctly account for the manner in which the $g_{H_iH_jH_k}$ are
defined in Ref.~\cite{CPsuperH}.  The Higgs propagator matrix is given
to a good approximation by~\cite{widths,Ellis2006}
\begin{eqnarray}
      iD(M^2) &=& \nn \\ 
         & & \hspace{-.6in}
           i\left(\begin{array}{ccc}
	    M^2 - m_{\sss H_1}^2 + i \, {\rm Im}\widehat{\Pi}_{11} & 
	     i \, {\rm Im}\widehat{\Pi}_{12} & i \, {\rm Im}\widehat{\Pi}_{13} \\
            i \, {\rm Im}\widehat{\Pi}_{21} & 
	     M^2 - m_{\sss H_2}^2 + i \, {\rm Im}\widehat{\Pi}_{22} & 
	     i \, {\rm Im}\widehat{\Pi}_{23} \\
            i \, {\rm Im}\widehat{\Pi}_{31} & i \, {\rm Im}\widehat{\Pi}_{32} &
	     M^2 - m_{\sss H_3}^2 + i \, {\rm Im}\widehat{\Pi}_{33} \\
	    \end{array}\right)^{-1} \; .
	   \label{eq:Higgsprop}
\end{eqnarray}
Expressions for the absorptive parts of the Higgs-boson self-energies,
${\rm Im}\widehat{\Pi}_{ij}(M^2)$, may be found in Ref.~\cite{widths}.
The Appendix of the present work contains a brief discussion of the
various couplings $g_{\sss H_i\tilde{t}_2^*\tilde{t}_1}$, $g_{\sss
H_jH_1H_1}$, $g_{\sss H_jVV}$ and $g_{\sss H_1H_jZ}$. Of these
couplings, only $g_{\sss H_i\tilde{t}_2^*\tilde{t}_1}$ is complex.  We
adopt the notation of Ref.~\cite{CPsuperH} for these couplings, except
in the case of $g_{\sss H_1H_jZ}$.

To obtain the width for ${\tilde t}_2^- \to {\tilde t}_1^- f_1f_2$, we
multiply the respective amplitude by its complex conjugate, sum over
the polarization states of the vector boson(s) in the final state (if
appropriate) and integrate over the squares of the two invariant
masses, $M^2$ and $\rho^2$, to obtain
\begin{eqnarray}
  \Gamma_{f_1f_2} = \frac{S_F}{256 \pi^3 m_{\tilde{t}_2}^3}\int
        \left(\sum_{\mbox{\scriptsize{pol'ns}}}\left|{\cal A}_{f_1f_2}\right|^2\right)
        dM^2 d\rho^2 \; ,
	\label{eq:Gamma}
\end{eqnarray}
where $S_F=1$ for $f_1f_2=WW,ZH_1$ and $S_F=\frac{1}{2}$ for
$f_1f_2=H_1H_1,ZZ$.  For the $H_1H_1$ case there is no sum over
polarizations and the calculation is relatively straightforward.  For
the $ZZ$ and $WW$ cases there are two sums over polarization states
(one for each vector particle in the final state), leading to ten
separate terms, each with its own kinematical factor.  The terms are
proportional to $\left|B\right|^2$, $\mbox{Re}\left(BC^*\right)$, etc.
The $ZH_1$ final state only requires one sum over polarization states,
resulting in three separate terms.  The appropriate range of
integration for $M^2$ and $\rho^2$ may be found, for example, in
Ref.~\cite{pdg}.

The width $\overline{\Gamma}_{f_1f_2}$ can be similarly defined for
the CP-conjugate process ${\tilde t}_2^+ \to {\tilde t}_1^+ {\bar
f_1}{\bar f_2}$.\footnote{Since the Higgs bosons are typically
mixtures of scalar and pseudoscalar states when CP is broken, some of
the final states that we consider are not eigenstates of CP.
Nevertheless, the widths we calculate for what we are calling the
``CP-conjugate'' processes are in fact the widths that would be
measured experimentally.  Furthermore, the asymmetries that we
calculate are zero in the CP-even limit.}  The only difference
compared to the expression in Eq.~(\ref{eq:Gamma}) is that one must
complex conjugate the weak phases in $B_{f_1f_2},\ldots,E_{f_1f_2}$,
which amounts to making the replacements $g_{\sss
H_i\tilde{t}_j^*\tilde{t}_k}\to g_{\sss
H_i\tilde{t}_j^*\tilde{t}_k}^*$ and $U^{\tilde t}\leftrightarrow
U^{{\tilde t}*}$ in the various expressions, as well as $i\to-i$ in
Eq.~(\ref{eq:bzh}).  Some further discussion may be found in the
Appendix.  ($U^{\tilde t}$ is the stop mixing matrix.)  For the rate
asymmetries to be non-zero, we require (at least) two interfering
amplitudes, and these must have a non-zero relative strong phase as
well as a non-zero relative weak phase.  The weak phases appear in the
various couplings and the strong phases are provided by absorptive
pieces in the Higgs propagator matrix.  (In principle, there are also
strong phases associated with the widths of the $Z$ and the squarks in
the diagrams in Fig.~\ref{fig:feyn_diag2}.  The effects of these
widths are suppressed, however, since the associated (s)particles are
always off-shell in our calculation.)

The rate asymmetries for the four cases are then defined to be
\bea
     A_{\mbox{\scriptsize CP}}({\tilde t}_2 \to {\tilde t}_1 f_1f_2) 
      = \frac{\Gamma_{f_1f_2}-\overline{\Gamma}_{f_1f_2}}{\Gamma_{f_1f_2}+\overline{\Gamma}_{f_1f_2}} \; .
\eea
Our previous paper~\cite{SUSYCP} contains an extended discussion of
the behaviour of CP asymmetries as functions of the invariant mass
$M$.  The analysis in the present case is complicated by the
non-negligible contributions of diagrams that are not in the
$s$-channel.  Nevertheless, one could still define differential widths
as functions of $M$, and these would still be expected to exhibit
resonant peaks near $M\approx m_{H_{2,3}}$.

\begin{figure}[t]
\begin{center}
\resizebox{5.5in}{!}{\includegraphics*{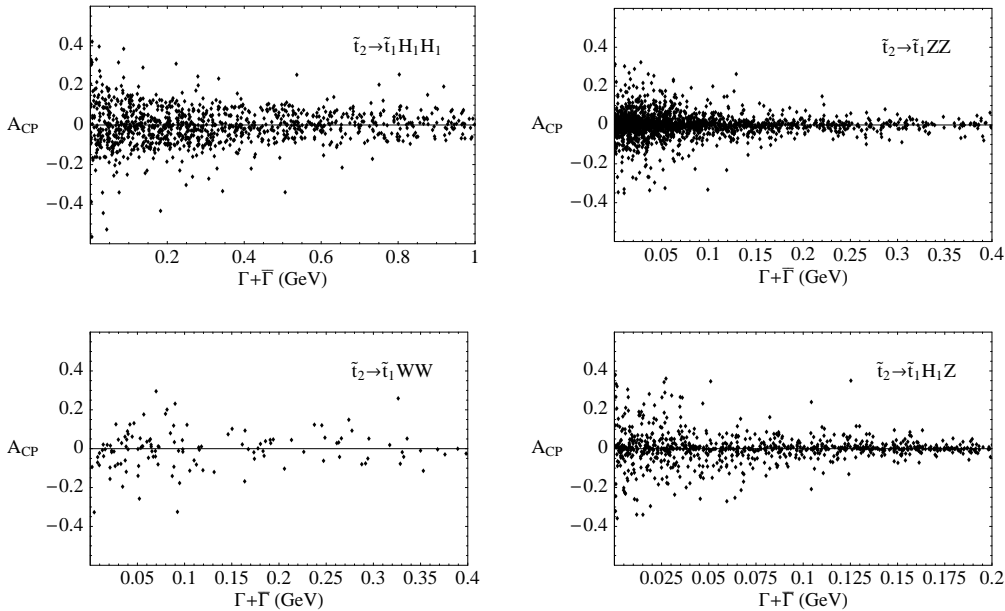}}
\caption{Scatter plots showing rate asymmetries for $\tilde{t}_2\to
\tilde{t}_1 H_1H_1, \tilde{t}_1ZZ, \tilde{t}_1W^+W^-$ and
$\tilde{t}_1 Z H_1$.  The horizontal axis in each plot 
gives the sum of the widths for the process and the anti-process for
the decay in question.}
\label{fig:stop_hh_zz_ww_hz}
\end{center}
\end{figure}

We now turn to a numerical investigation of rate asymmetries for
$\tilde{t}_2\to \tilde{t}_1 H_1H_1, \tilde{t}_1ZZ, \tilde{t}_1W^+W^-$
and $\tilde{t}_1ZH_1$.  In our numerical work we have made extensive
use of the computer program {\tt CPsuperH}~\cite{CPsuperH}.
Note that since we are using the widths of the Higgs
bosons to provide the required strong phase, the rate asymmetries will
only be non-negligible if there is a significant overlap of the
interfering resonances (see Ref.~\cite{SUSYCP} for further discussion
of this point). Fortunately, as noted above, it is not uncommon to
have $m_{H_2}\approx m_{H_3}$ in SUSY.

Figure~\ref{fig:stop_hh_zz_ww_hz} shows several scatter plots of the
rate asymmetries for $\tilde{t}_2\to \tilde{t}_1 H_1H_1,
\tilde{t}_1ZZ, \tilde{t}_1W^+W^-$ and $\tilde{t}_1ZH_1$. In these
plots we have allowed several SUSY parameters to vary in specified
ranges, taking $m_{H^\pm}\in(160-500)$~GeV, $\tan\beta\in (1-15)$,
$\mu\in (200-1400)$~GeV (without loss of generality, the
$\mu$ parameter is taken to be real and positive -- see Ref.~\cite{SUSYCP} for
further discussion), $m_{\tilde{Q}_3},m_{\tilde{U}_3},m_{\tilde{D}_3}
\in(300-700)$~GeV, $\left|A_t\right|\in(400-2000)$~GeV and
$\phi_{A_t}\in(0^\circ-360^\circ)$. All other input parameters have
been assigned fixed values and have been taken to be real.\footnote{
For completeness, we list here some of the other parameter choices, in 
{\tt CPsuperH} notation: $M_1=100$~GeV, $M_2=200$~GeV, 
$m_{\tilde{g}}=M_3=1000$~GeV,
$m_{\tilde{L}_3}=150$~GeV, $m_{\tilde{E}_3}=600$~GeV,
$A_b=1000$~GeV and $A_\tau=750$~GeV.  With these parameter choices,
and after applying cuts noted subsequently in the text, 
some of the supersymmetric particles have the following
mass ranges: $m_{\tilde{\nu}_\tau}\in(136-138)$~GeV,
$m_{\tilde{t}_1}\in(199-558)$~GeV,
$m_{\tilde{t}_2}\in(468-883)$~GeV,
$m_{\tilde{b}_1}\in(255-685)$~GeV,
$m_{\tilde{\tau}_1}\in(146-156)$~GeV,
$m_{\tilde{\chi}_1^\pm}\in(196-200)$~GeV and
$m_{\tilde{\chi}_1^0}\in(100-101)$~GeV.  Also, the mass ranges
for the three Higgs bosons are
$m_{H_1}\in(100-126)$~GeV, 
$m_{H_2}\in(161-490)$~GeV
and $m_{H_3}\in(161-495)$~GeV.}
 We have
insisted that the mass of the lightest Higgs boson be greater than or
equal to 100~GeV and have chosen values for the charged Higgs
mass and $\tan\beta$ that are consistent with the recent bound from
Belle, $\tan\beta/m_{\sss H^\pm}\lsim 0.146~\mbox{GeV}^{-1}$
\cite{belle}.  We have also insisted that all supersymmetric particles
have masses greater than or equal to 100~GeV and that the lighter stop be
kinematically allowed to decay by at least one of the following
two modes: $\tilde{t}_1\to t \tilde{\chi}_1^0$ or 
$\tilde{t}_1^\pm\to b\tilde{\chi}_1^\pm$.
For the purpose of these plots we have set the
widths of intermediate stops and sbottoms to be
10~GeV~\cite{stopwidth}.  (Varying these widths does not seem to have
a significant effect on the numerical values obtained for the
asymmetries, as long as the sbottoms are not allowed to go on shell.)
The horizontal axes in these plots show the sum of the widths for the
process and the anti-process.  If we assume a typical stop width to be
approximately 10~GeV, we see that it is possible to have large rate
asymmetries while simultaneously having relatively large branching
ratios. For example, in $\tilde{t}_2\to \tilde{t}_1 H_1H_1$ it is
possible to have an asymmetry with magnitude of order $20$$-$$30\%$ when
$\Gamma_{H_1H_1}+\overline{\Gamma}_{H_1H_1}\simeq 0.2$~GeV. Among the decays
considered, $\tilde{t}_2\to \tilde{t}_1 H_1H_1$ is clearly favoured in
that it tends to have a larger width for a given CP
asymmetry. Experimentally, however, some of the other decay channels
may be favoured due to increased detection efficiencies.  Note that
the $WW$ final state has fewer data points compared to the other plots
because parameter sets for which a sbottom would have gone on shell
have been discarded for that plot.

In Figure~\ref{fig:stop_hh_zz_ww_hz} we have allowed the phase of the
trilinear coupling $A_t$ to vary, but have set all other phases to
zero.  In addition to the results shown in
Fig.~\ref{fig:stop_hh_zz_ww_hz}, we have also performed a limited
analysis of the effects on the CP asymmetries due to the phases of
other SUSY parameters.  The strongest effects come from $\phi_{A_b}$
and $\phi_{M_3}$, the phases of the trilinear coupling $A_b$ and
gluino mass $M_3$, respectively.  The phase $\phi_{A_b}$ can affect
the asymmetries due to the involvement of $A_b$ in the mixing of the
Higgs bosons.  The phase of $M_3$ comes into play through its effect
on the Higgs-bottom-bottom effective vertex\footnote{Even though the
decays under consideration do not involve this effective vertex in a
direct way, the strength of this vertex can affect the widths of the
two heavier Higgs bosons, whose values can undergo large variations
for certain values of $\phi_{M_3}$. This could lead, for instance, to
a larger overlap between the two heavier Higgs bosons in the available phase
space, giving rise to an enhancement of the CP asymmetries.}.  The
phase $\phi_{A_b}$ only seems to produce a non-negligible effect when
$| A_b |$ is greater than 5~TeV (due to the difference in the top and
bottom Yukawa couplings).  Allowing $\phi_{M_3}$ to assume a nonzero
value can lead to changes in the CP asymmetries of order $\pm 0.1$ if
$\left|M_3\right|$ is taken to be approximately 1~TeV.

To summarize, many physicists believe that supersymmetry (SUSY) is the
physics that lies beyond the standard model, and that it will be found
at a future high-energy collider. Assuming that this is the case, we
will want to measure the CP-violating SUSY parameters. With this in
mind, in a previous paper \cite{SUSYCP}, we studied the decay ${\tilde
t}_2 \to {\tilde t}_1 \tau^- \tau^+$. We found that the CP asymmetries
could be large, but that they require the measurement of the spin of
one or both of the final-state $\tau$'s.

In the present paper we have studied the decays ${\tilde t}_2 \to
{\tilde t}_1 HH$, ${\tilde t}_2 \to {\tilde t}_1 ZZ$, ${\tilde t}_2
\to {\tilde t}_1 W^+ W^-$ and ${\tilde t}_2 \to {\tilde t}_1 ZH$,
constructing rate asymmetries that are sensitive to CP violation in
the underlying SUSY theory.  There are several ways in which the
current work complements that performed in our previous study.  First,
as we have shown, the CP asymmetries considered here do not require
the measurement of any spins -- the rate asymmetries alone are
measurable and they can be relatively large.  Second, we had found in
our previous work that ${\tilde t}_2 \to {\tilde t}_1 \tau^- \tau^+$
is sensitive primarily to $\phi_{\sss A_t}$, the phase of the
trilinear coupling $A_t$.  As noted above, the decays considered in
this work are also sensitive primarily to $\phi_{A_t}$, although the
phases associated with the trilinear coupling $A_b$ and the gluino
mass $M_3$ can also affect the asymmetries.  Third, it is worth
noting that the decays considered here tend to involve a different
range of Higgs masses than those considered in our previous work.  For
example, in the present case we need to have $m_{H_{2,3}}\gsim 2 m_W$
in order for ${\tilde t}_2 \to {\tilde t}_1 W^+ W^-$ to yield an
appreciable asymmetry along with a moderately large branching
fraction.  In our former work, non-negligible asymmetries 
and branching fractions could be obtained for lower Higgs masses.

Of the processes studied in this work, we find that ${\tilde t}_2 \to
{\tilde t}_1 HH$ tends to have the largest CP asymmetry and width, and
so it is the process which is perhaps most amenable to study. (Note
that experimental detection efficiencies may make other decay modes
more favourable.)  As we have emphasized, the decays of stops depend
primarily on $\phi_{\sss A_t}$, and so do not provide a sensitivity to
a large number of phases. It would be of considerable interest to
investigate CP asymmetries in analogous decays of sbottoms or staus to
assess their sensitivity to various SUSY phases ($\phi_{A_b}$,
$\phi_{A_\tau}$, and probably also $\phi_{A_t}$).  We are currently
beginning such a study to determine if such decays might provide
complementary tools for the study of CP violation in SUSY.


\begin{acknowledgments}
K.K. thanks the Physics Department at the University of Montr\'{e}al
for its hospitality during the initial stages of this work.  We also
thank Georges Azuelos and Makiko Nagashima for helpful discussions;
and J.S. Lee for helpful correspondence and for giving us a subroutine
for calculating the Higgs propagator matrix. This work was financially
supported by NSERC of Canada. The work of K.K. was supported in part
by the U.S.\ National Science Foundation under Grants PHY--0301964 and
PHY--0601103.
\end{acknowledgments}


\appendix*


\section{Expressions for Coupling Constants and Amplitudes}


In this Appendix we describe some of the Higgs couplings in
Eqs.~(\ref{eq:bhh})-(\ref{eq:bzh}) in more detail.  We also provide
the full expressions for the amplitudes referred to in
Eqs.~(\ref{eq:ahh})-(\ref{eq:azh}).

The couplings between the stops and Higgs bosons are defined
as follows~\cite{CPsuperH},
\beq
   vg_{\sss H_i\tilde{t}_j^*\tilde{t}_k} = 
      \left(\Gamma^{\alpha\tilde{t}^*\tilde{t}}\right)_{\beta\gamma} 
      O_{\alpha i} U^{\tilde{t}*}_{\beta j} U^{\tilde{t}}_{\gamma k} \; ,
\eeq
where $O$ and $U^{\tilde{t}}$ denote the Higgs and stop mixing
matrices~\cite{CPsuperH}, respectively, and where $i=1,2,3$ and
$j,k=1,2$.  As usual, $v=\sqrt{v_1^2+v_2^2}$ is defined in terms of
the vacuum expectation values of the two Higgs doublets.  Expressions
for the $2\times 2$ matrices $\Gamma^{\alpha\tilde{t}^*\tilde{t}}$ may
be found in Appendix B of Ref.~\cite{CPsuperH}.  These matrices depend
on the SUSY parameters $A_t$, $\mu$, $\cos\beta$ and $\sin\beta$
(where $\tan\beta\equiv v_2/v_1$).  It is interesting to note that the
couplings involving scalar Higgs bosons are real and those involving
the pseudoscalar Higgs are purely imaginary.  These
couplings are generally complex if CP is broken.

For the Higgs-Higgs-$Z$ vertex we adopt the same notation as in
Ref.~\cite{cepw} (this differs somewhat from that employed in
Ref.~\cite{CPsuperH}):
\begin{eqnarray}
  {\cal L}_{HHZ} = \frac{g}{2\cos\theta_W}\sum_{j>i} g_{H_iH_jZ}Z^\mu
     H_i \stackrel{\leftrightarrow}{\partial}_\mu H_j \; ,
\end{eqnarray}
where
\begin{eqnarray}
  g_{H_iH_jZ} = \left\{\begin{array}{ll}
           O_{3i}\left(c_\beta O_{2j}-s_\beta O_{1j}\right)
	   - O_{3j}\left(c_\beta O_{2i}-s_\beta O_{1i}
	   \right), & j>i \nonumber\\
	   0, & \mbox{otherwise,} \\
	   \end{array}\right.
\end{eqnarray}
with $c_\beta=\cos\beta$ and $s_\beta=\sin\beta$.  The Higgs-$W$-$W$
and Higgs-$Z$-$Z$ vertices are both proportional to the coupling
$g_{H_iVV}$, defined as follows~\cite{CPsuperH}:
\begin{eqnarray}
  g_{H_iVV} = c_\beta O_{1i}+s_\beta O_{2i} \; .
\end{eqnarray}
Finally, expressions for the trilinear Higgs couplings $g_{H_iH_jH_k}$ are
given in Ref.~\cite{CPsuperH} and references therein.

We now turn to a consideration of the full expressions for the
amplitudes in Eqs.~(\ref{eq:ahh})-(\ref{eq:azh}).  The parameters
$B_{f_1f_2}$ always include the Higgs-mediated pieces, but also
include other contributions with similar kinematical structures.  In
the text these have been parameterized as
\begin{eqnarray}
  B_{f_1f_2} = B_{f_1f_2}^{\mbox{\scriptsize{Higgs}}}+\delta B_{f_1f_2} \; .
    \nonumber
\end{eqnarray}
The expressions for the Higgs-mediated pieces,
$B_{f_1f_2}^{\mbox{\scriptsize{Higgs}}}$, are given in
Eqs.~(\ref{eq:bhh})-(\ref{eq:bzh}).  In the expressions below we
employ the Breit-Wigner form of the propagator for sbottom- and
stop-mediated graphs, defining
\begin{eqnarray}
  i\widetilde{D}\left(p^2,m^2,\Gamma\right) & \equiv & \frac{i}{p^2-m^2+i\Gamma m} \; ,
\end{eqnarray}
where we have used a tilde to distinguish the Breit-Wigner propagator
from the $3\times 3$ Higgs propagator matrix defined in
Eq.~(\ref{eq:Higgsprop}).  We also use a Breit-Wigner-type propagator
for $Z$-mediated graphs.  Working in Unitary gauge, we take the
propagator to be $i\widetilde{D}\left(p_Z^2,m_Z^2,\Gamma_Z\right)
\left(-g^{\alpha\beta}+p_Z^\alpha p_Z^\beta/m_Z^2\right)$.  We employ
the following definitions for the various combinations of momenta that
appear in the propagators,
\begin{eqnarray}
  M^2 & = & \left(p_1+p_2\right)^2 \; , \nonumber \\
  \rho^2 & = & \left(p_1+p_{\tilde{t}_1}\right)^2 \nonumber\; , \\
  \xi^2 & = & \left(p_2+p_{\tilde{t}_1}\right)^2 \nonumber\; ,
\end{eqnarray}
where we note that $\xi^2$ can be written as a function of $M^2$,
$\rho^2$ and various masses.

The Feynman rules used to build many of the following amplitudes were
extracted from Ref.~\cite{Rosiek} (with appropriate modifications to
allow for scalar-pseudoscalar mixing among the Higgs bosons).  Amplitudes
corresponding to the charge conjugated decays can be obtained from
those listed below [and in Eqs.~(\ref{eq:bhh})-(\ref{eq:bzh})] by taking
the complex conjugate everywhere except in the propagator functions
$D_{ij}$ and $\widetilde{D}$.  This is
equivalent to making the replacements $g_{\sss
H_i\tilde{t}_j^*\tilde{t}_k}\to g_{\sss
H_i\tilde{t}_j^*\tilde{t}_k}^*$ and $U^{\tilde t}\leftrightarrow
U^{{\tilde t}*}$ in the various expressions, as well as $i\to-i$ in
Eq.~(\ref{eq:bzh}).


\subsection{\boldmath ${\tilde t}_2^- \to {\tilde t}_1^- H_1H_1$}


The amplitude for ${\tilde t}_2 \to {\tilde t}_1 H_1H_1$ is given by
\begin{eqnarray}
  B_{HH} & = & B_{HH}^{\mbox{\scriptsize{Higgs}}} 
              -v^2\sum_j g_{\sss H_1\tilde{t}_2^*\tilde{t}_j}
              g_{\sss H_1\tilde{t}_j^*\tilde{t}_1}\left[
	       \widetilde{D}\left(\rho^2,m_{\tilde{t}_j}^2,\Gamma_{\tilde{t}_j}\right)+
	       \widetilde{D}\left(\xi^2,m_{\tilde{t}_j}^2,\Gamma_{\tilde{t}_j}\right)
	        \right] \nonumber \\
	 & &  -\left(1-\frac{8}{3}\sin^2\theta_W\right)
	       \left[\left(O_{11}\right)^2-\left(O_{21}\right)^2+\left(O_{31}\right)^2
		 \left(s_\beta^2-c_\beta^2\right)\right]\eta_{HH}\; ,
\end{eqnarray}
where
\begin{eqnarray}
  \eta_{HH} = \frac{g^2}{4 \cos^2\theta_W} U^{\tilde{t}*}_{12} U^{\tilde{t}}_{11} \; .
\end{eqnarray}
%


\subsection{\boldmath ${\tilde t}_2^- \to {\tilde t}_1^- ZZ$}


The amplitudes for ${\tilde t}_2 \to {\tilde t}_1 ZZ$ are as follows:
\begin{eqnarray}
  B_{ZZ} & = & B_{ZZ}^{\mbox{\scriptsize{Higgs}}}+
      \left(2-\frac{16}{3}\sin^2\theta_W\right)\eta_{ZZ} \; , \\
  C_{ZZ} & = & -4\sum_j
    \left(\left|U^{\tilde{t}}_{1j}\right|^2-\frac{4}{3}\sin^2\theta_W\right)
      \widetilde{D}\left(\xi^2,
            m_{\tilde{t}_j}^2,\Gamma_{\tilde{t}_j}\right)\eta_{ZZ} \; , \\
  D_{ZZ} & = & -4\sum_j
    \left(\left|U^{\tilde{t}}_{1j}\right|^2-\frac{4}{3}\sin^2\theta_W\right)
      \widetilde{D}\left(\rho^2,
            m_{\tilde{t}_j}^2,\Gamma_{\tilde{t}_j}\right)\eta_{ZZ} \; , \\
  E_{ZZ} & = & C_{ZZ}+D_{ZZ} \; .
\end{eqnarray}
Each of the non-Higgs-mediated pieces is proportional to the same
complex quantity $\eta_{ZZ}$:
\begin{eqnarray}
  \eta_{ZZ} = \eta_{HH} = \frac{g^2}{4 \cos^2\theta_W} U^{\tilde{t}*}_{12} U^{\tilde{t}}_{11} \; .
\end{eqnarray}
%


\subsection{\boldmath ${\tilde t}_2^- \to {\tilde t}_1^- W^+ W^-$}


The amplitudes for ${\tilde t}_2 \to {\tilde t}_1 W^+ W^-$ are given by,
\begin{eqnarray}
  B_{WW} & = & B_{WW}^{\mbox{\scriptsize{Higgs}}}+
      \left[1+\left(\rho^2-\xi^2\right)
	   \widetilde{D}\left(M^2,m_Z^2,\Gamma_Z\right)\right]\eta_{WW} \; ,\\
  C_{WW} & = & 4\, \widetilde{D}\left(M^2,m_Z^2,\Gamma_Z\right)\eta_{WW}\; , \\
  D_{WW} & = & -4 \left[\widetilde{D}\left(M^2,m_Z^2,\Gamma_Z\right) + 
        \sum_j \left| U^{\tilde{b}}_{1j}\right|^2\left|V_{tb}\right|^2
	    \widetilde{D}\left(\rho^2,m_{\tilde{b}_j}^2,\Gamma_{\tilde{b}_j}\right)
	      \right]\eta_{WW} \; , \\
  E_{WW} & = & C_{WW}+D_{WW} = -4 \sum_j \left| U^{\tilde{b}}_{1j}\right|^2\left|V_{tb}\right|^2
            \widetilde{D}\left(\rho^2,m_{\tilde{b}_j}^2,\Gamma_{\tilde{b}_j}\right)
	      \eta_{WW} \; .
\end{eqnarray}
In the numerical work we set $V_{tb}=1$. Note that each of the
non-Higgs terms is proportional to the same complex quantity
$\eta_{WW}$, which is defined as follows:
\begin{eqnarray}
  \eta_{WW} = \frac{g^2}{2} U^{\tilde{t}*}_{12} U^{\tilde{t}}_{11} \; .
\end{eqnarray}
%


\subsection{\boldmath ${\tilde t}_2^- \to {\tilde t}_1^- ZH_1$}


The amplitudes for ${\tilde t}_2 \to {\tilde t}_1 ZH_1$ are as
follows:
\begin{eqnarray}
  B_{ZH} & = & B_{ZH}^{\mbox{\scriptsize{Higgs}}}
       -\frac{2}{m_Z}\left(m_{\tilde{t}_2}^2-m_{\tilde{t}_1}^2-m_Z^2\right)
       \widetilde{D}\left(M^2,m_Z^2,\Gamma_Z\right)g_{H_1VV}\, \eta_{ZH} \nonumber\\
      & & -\frac{vg}{\cos\theta_W}\sum_j\left(U^{\tilde{t}*}_{12} U^{\tilde{t}}_{1j}-
      \frac{4}{3}\sin^2\theta_W\delta^{2j}\right) 
            \widetilde{D}\left(\xi^2,
            m_{\tilde{t}_j}^2,\Gamma_{\tilde{t}_j}\right)
	    g_{\sss H_1\tilde{t}_j^*\tilde{t}_1} \; ,\\
  C_{ZH} & = & 4\,m_Z\widetilde{D}\left(M^2,m_Z^2,\Gamma_Z\right)
               g_{H_1VV}\, \eta_{ZH} \nonumber\\
      & & -\frac{vg}{\cos\theta_W}\sum_j\left(U^{\tilde{t}*}_{12} U^{\tilde{t}}_{1j}-
      \frac{4}{3}\sin^2\theta_W\delta^{2j}\right) 
            \widetilde{D}\left(\xi^2,
            m_{\tilde{t}_j}^2,\Gamma_{\tilde{t}_j}\right)
	    g_{\sss H_1\tilde{t}_j^*\tilde{t}_1} \nonumber \\
      & & -\frac{vg}{\cos\theta_W}\sum_j\left(U^{\tilde{t}*}_{1j} U^{\tilde{t}}_{11}-
      \frac{4}{3}\sin^2\theta_W\delta^{1j}\right) 
            \widetilde{D}\left(\rho^2,
            m_{\tilde{t}_j}^2,\Gamma_{\tilde{t}_j}\right)
	    g_{\sss H_1\tilde{t}_2^*\tilde{t}_j} \; ,
\end{eqnarray}
where
\begin{eqnarray}
  \eta_{ZH} = \eta_{HH} = \frac{g^2}{4 \cos^2\theta_W} U^{\tilde{t}*}_{12} U^{\tilde{t}}_{11} \; .
\end{eqnarray}
%


\end{document}